\documentclass[aps, prl,a4paper,preprint ]{revtex4-1}

\usepackage[latin1]{inputenc}
\usepackage{graphicx}
\usepackage{amsmath}
\usepackage{amsthm}
\usepackage{amssymb}

\def\norm#1{\left\|#1\right\|}

\def\qquadtext#1{{\qquad\text{#1}\qquad}}
\def\qquadand{\qquadtext{and}}
\def\quadtext#1{{\quad\text{#1}\quad}}

\def\quadand{\quadtext{and}}

\def\dfrac#1#2{{\frac{d #1}{d #2}}}
\def\dfracslash#1#2{{d #1/d #2}}

\def\VE{{\boldsymbol E}}
\def\VB{{\boldsymbol B}}
\def\VX{{\boldsymbol X}}

\def\Vbeta{{\boldsymbol \beta}}
\def\Vx{{\boldsymbol x}}

\def\tauhat{\hat\tau}

\def\THat{\hat T}

\def\VEE{{\boldsymbol{E_0}}}

\def\VET{{\boldsymbol{E}}_{\textup{Tot}}}
\def\Va{{\boldsymbol a}}
\def\Vr{{\boldsymbol r}}
\def\Vrhat{{{\boldsymbol n}}}

\def\rhoLab{{\rho_{\textup{Lab}}}}
\def\tstraight{{t_\text{s}}}
\def\const{\frac{q}{4\pi\epsilon_0}}
\begin{document}

\begin{flushright}
Cockcroft-11-16
\end{flushright}
\vspace{3em}

\title{Bending a Beam to Significantly Reduce Wakefields of Short Bunches}
\author{Jonathan Gratus}
\thanks{j.gratus@lancs.ac.uk}
\author{Michael R. Ferris}
\thanks{m.ferris@lancs.ac.uk}
\affiliation{Department of Physics, Lancaster University, Lancaster, Lancashire, UK,  LA1 4YB}
\affiliation{The Cockcroft Institute,
Daresbury Laboratory,
Daresbury Science and Innovation Campus, Keckwick Lane, Daresbury, Warrington, Cheshire, UK, WA4 4AD}

\begin{abstract}
A method of significantly reducing wakefields generated at
collimators is proposed, in which the path of a beam is slightly bent
before collimation. This is applicable for short bunches and can
reduce the wakefields by a factor of around 7 for present day
free electron lasers and future colliders.
\end{abstract}
\maketitle

Electromagnetic wakefields are created when an accelerated bunch of
charged particles passes a discontinuity in the metallic structure of
the beam pipe. Fields caused by geometric discontinuities, for example in cavities and collimators, are known as geometric wakefields, and can
induce instabilities and emittance growth in the particle beam. As a
result there is much interest in methods for reducing the geometric
wakefields produced by a charged bunch of particles passing through a
collimator. The customary approach is to reduce the taper angle of the
collimator. Early work on the
calculation of wakefields from smoothly tapered structures was
pioneered by Yokoya \cite{Yokoya90}, Warnock \cite{Warnock93} and
Stupakov \cite{Stupakov96, Stupakov01}. More recent investigations by Stupakov,
  Bane and Zagorodnov \cite{Stupakov07, Bane07, Bane10} and 
Podobedov and Krinsky \cite{Podobedov06, Podobedov07} have also looked at the effect of altering the transverse cross section of the
 collimator. A detailed analysis of the
numerical and analytic calculation of collimator wakefields, including
an informative introduction to the topic, may be found in \cite{Smith11}.

In this article an alternative approach is suggested whereby
geometric wakefields are reduced by altering the path of the beam
prior to collimation. This approach is facilitated by the highly relativistic regime in which
lepton accelerators operate, where as shown in the following,
the Coulomb field given from the Li\'{e}nard-Wiechert potential is highly collimated in the direction
of motion. It turns out the standard \emph{pancake} field associated with a
highly relativistic particle takes a finite time to develop to a given
width. Thus by placing the collimator sufficiently close to a
bending dipole the radius of the pancake remains smaller than the
width of the aperture of the collimator. Using this method reduction of wakefields by factors of around 7
are feasible for some present day energies and bunch lengths.

For a particle of charge $q$ undergoing arbitrary motion $\Vx(\tau)$, where
$\tau$ is the particle's proper time, the Li\'{e}nard-Wiechert fields
at point $\VX$ and time $T$ are given \cite{Jackson99} by
%[
\begin{align}
\VE(\VX,T)\!=\!\VE_{\text{LW}}\Big(\!\VX\!-\!\Vx(\tau_R),\Vbeta(\tau_R),\Va(\tau_R)\!\Big)
\label{EB_TX}
\end{align}
%]
and
%[
\begin{align}
\VB(\VX, T)\!=\!\VB_{\text{LW}}\Big(\!\VX\!-\!\Vx(\tau_R),\Vbeta(\tau_R),\Va(\tau_R)\!\Big).
\label{BB_TX}
\end{align}
%]
Here $\tau_R$ is the retarded proper time and $\Vx(\tau_R)$ the position of
the charged particle at the retarded proper time. The $3$-vectors
%[
\begin{align*}
\Vbeta(\tau)\!=\!\frac{1}{c\gamma}\dfrac{\Vx}{\tau},
\qquadand
\Va(\tau)\!=\!\frac{1}{\gamma}\dfrac{\Vbeta}{\tau}
\end{align*}
%]
are the Newtonian velocity and acceleration divided by $c$, where
$\gamma\!=\!\sqrt{1+\norm{\dfracslash{\Vx}{\tau}}^2/c^2}$.
The functions $\VE_{\text{LW}}$ and $\VB_{\text{LW}}$ are given by
%[
\begin{align}
& \VE_{\text{LW}}(\!\Vr\!,\!\Vbeta\!,\!\Va\!)\!\!=\!\!\const\!\Bigg(\!
\frac{ (\Vrhat\!-\!\Vbeta)}{\gamma^2\! \norm{\Vr}^2\!(1\!-\!\Vbeta\cdot\Vrhat)^3}\!
+\!
\frac{ \Vrhat\times\Big(\!(\Vrhat\!-\!\Vbeta)\times\Va\!\Big)}
{c\!\norm{\Vr}\!(1\!-\!\Vbeta\cdot\Vrhat)^3}\!\Bigg)
\label{BB_FLW_E}
\end{align}
%]
and\vspace{-0.7cm}
%[
\begin{align}
\VB_{\text{LW}}(\Vr,\Vbeta,\Va)=\frac{1}{c}
\Vrhat\times\VE_{\text{LW}}(\Vr,\Vbeta,\Va),
\label{BB_FLW_B}
\end{align}
%]
where $\Vr=\VX-\Vx(\tau_R)$, $\Vrhat=\Vr/\norm{\Vr}$ and $\gamma=(1-\norm{\Vbeta}^2)^{-1/2}$.
 The first term in (\ref{BB_FLW_E}) will be referred to as the Coulomb field and the second term
as the Radiative field.

\begin{figure}[ht]
\begin{center}
\begin{tabular}{c@{\qquad\qquad}c}
\includegraphics[width=0.4\textwidth,viewport=22 293 553 515]{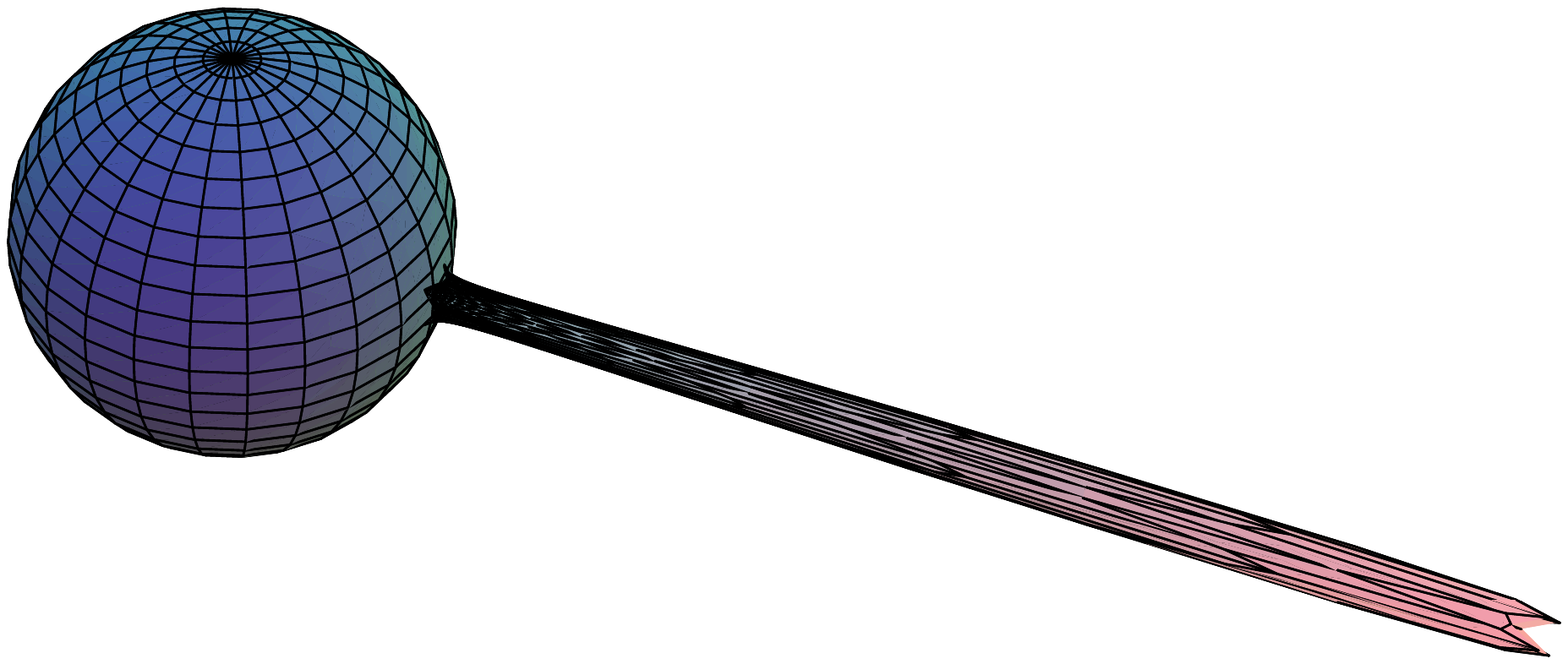}
&
\includegraphics[width=0.4\textwidth,viewport=25 306 540 500]{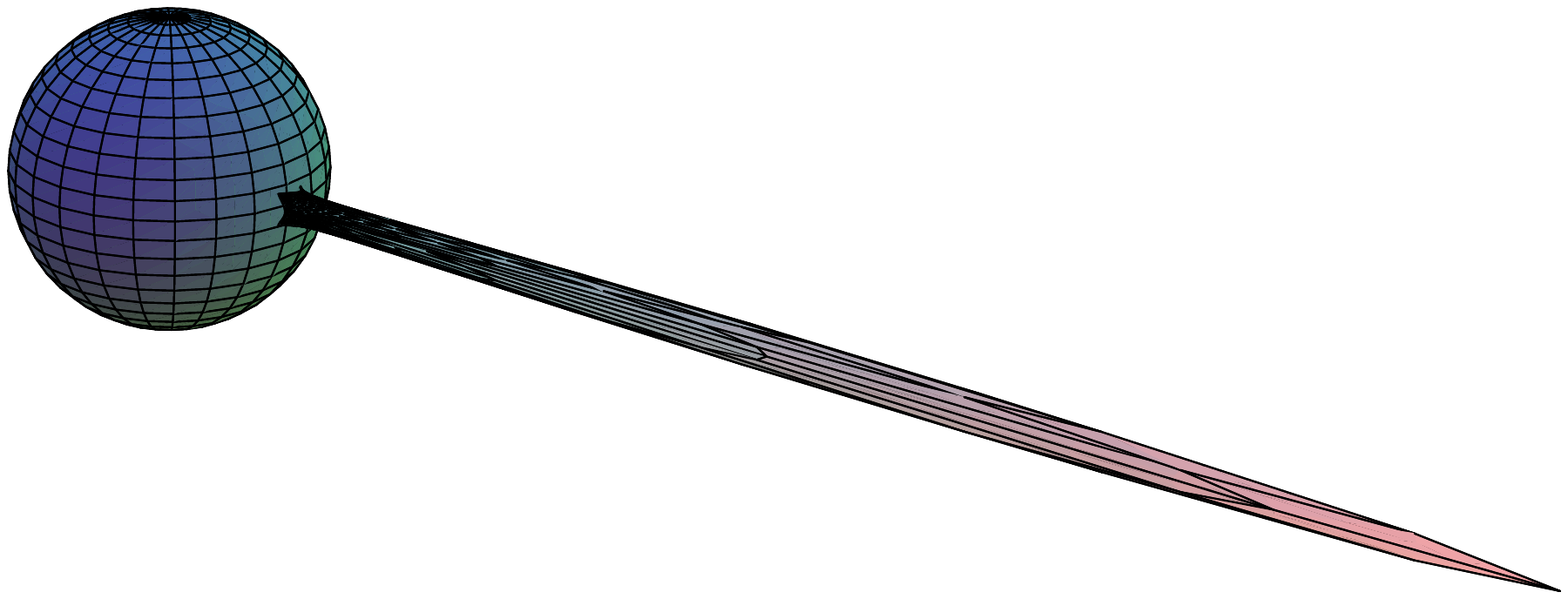}
\\
Coulomb field
&
Radiative field
\end{tabular}
\end{center}
\caption{The magnitude of the Coulomb and radiative fields for a high
  $\gamma$, given as height above the sphere. The bulk of the fields is
in the direction of motion.}
\label{fig_Fields}
\end{figure}

For particles moving close to the speed of light, i.e. with high
$\gamma$-factors and $\norm{\Vbeta}\approx 1$, the denominator in (\ref{BB_FLW_E}) is very small when $\Vrhat$ is in the direction of $\Vbeta$. Hence
$\VE_{\text{LW}}(\Vr,\Vbeta,\Va)$ is very large in the direction $\Vrhat\approx\Vbeta$ . FIG. \ref{fig_Fields} is a plot of
the magnitude of the Coulomb and Radiative fields for fixed $\norm{\Vr}$, $\Vbeta$ and $\Va$ as
a function of the spherical coordinates.

A relativistic particle undergoing nonlinear acceleration will generate a field primarily in
the instantaneous direction of motion of the the particle. This is true not just for the radiation field, where
it is usually described as a search light, but also for the Coulomb field. For high $\gamma$-factors the bulk of the fields, for both the
Coulomb and radiation terms, is inside an angle
$\Delta\phi\sim 1/\gamma$ where $\Delta\phi$ is the angle from the
direction of motion. By contrast, the field generated by a relativistic particle moving with constant velocity is flattened
towards the plane orthogonal to the direction of motion, and is often called a \emph{pancake} field.
 It is reasonable to ask how these two
radically different behaviours can be consistent.
\begin{figure}[ht]
\centerline{
\setlength{\unitlength}{0.08\textwidth}
\begin{picture}(10,2.5)
\put(0,0){\includegraphics[width=10\unitlength]{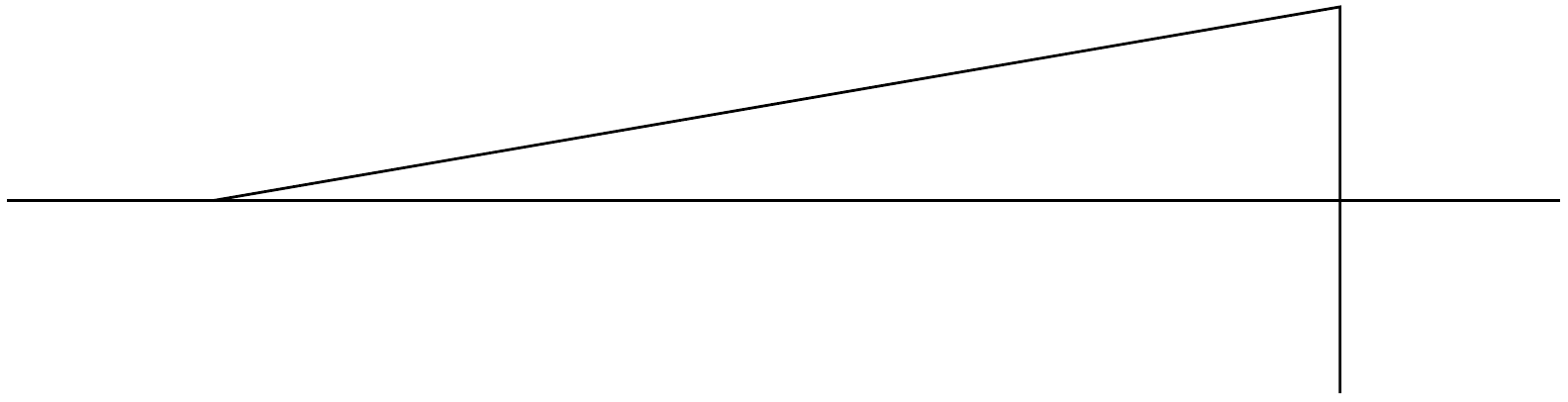}}
%\graphpaper[1](0,0)(10,2.5)
\put(8.6,1.5){$h$}
\put(5,0.7){$v\tstraight$}
\put(5,2.1){$c\tstraight$}
\put(8.6,2.5){R}
\put(8.6,.7){Q}
\put(1.0,.7){P}
\end{picture}
}
\caption{Showing the communication between a particle and its pancake}
\label{fig_Catchup}
\end{figure}
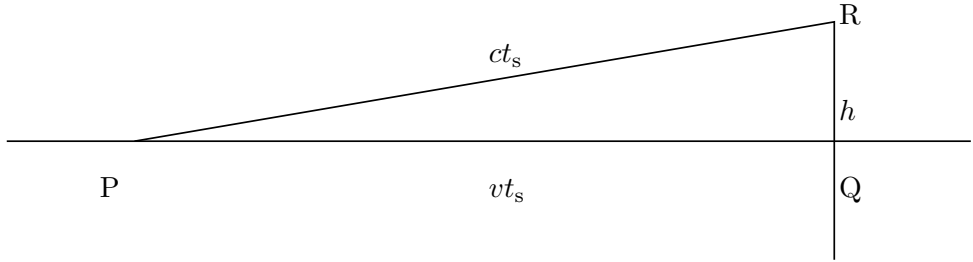

Consider a particle moving at velocity $v$ along the horizontal line $PQ$
in FIG. \ref{fig_Catchup}. Let $R$ be a point in the pancake a
distance $h$ from the particle, when the particle is at $Q$. The last
point at which the particle could communicate with the point $R$ is at
$P$, a length $v\tstraight$ from $Q$. Here $\tstraight$ is the time
it takes for light to travel from $P$ to $R$ and also the time for
the particle to travel from $P$ to $Q$. Then
$\norm{PR}=c\tstraight$ and $\norm{PQ}=v\tstraight$. Thus
$(c\tstraight)^2=h^2+(v\tstraight)^2$. Hence
$h^2=c^2\tstraight^2(1-v^2/c^2)=c^2\tstraight^2/\gamma^{2}$ so
$\tstraight=\gamma h/c$ and $\norm{PQ}=\gamma h v/c$. Thus a
particle needs to have travelled in a straight line for a length
$\norm{PQ}=\gamma h v/c$ in order for a pancake of radius $h$ to
develop. Looking at the fields which originate at $P$ and arrive at $R$, they
are at an angle approximately $\norm{RQ}/\norm{PR}=1/\gamma$. This is
consistent with FIG. \ref{fig_Fields}.

\begin{figure}[ht]
\centerline{
\setlength{\unitlength}{0.009\textwidth}
\begin{picture}(100,50)
\put(0,0){\includegraphics[width=100\unitlength]{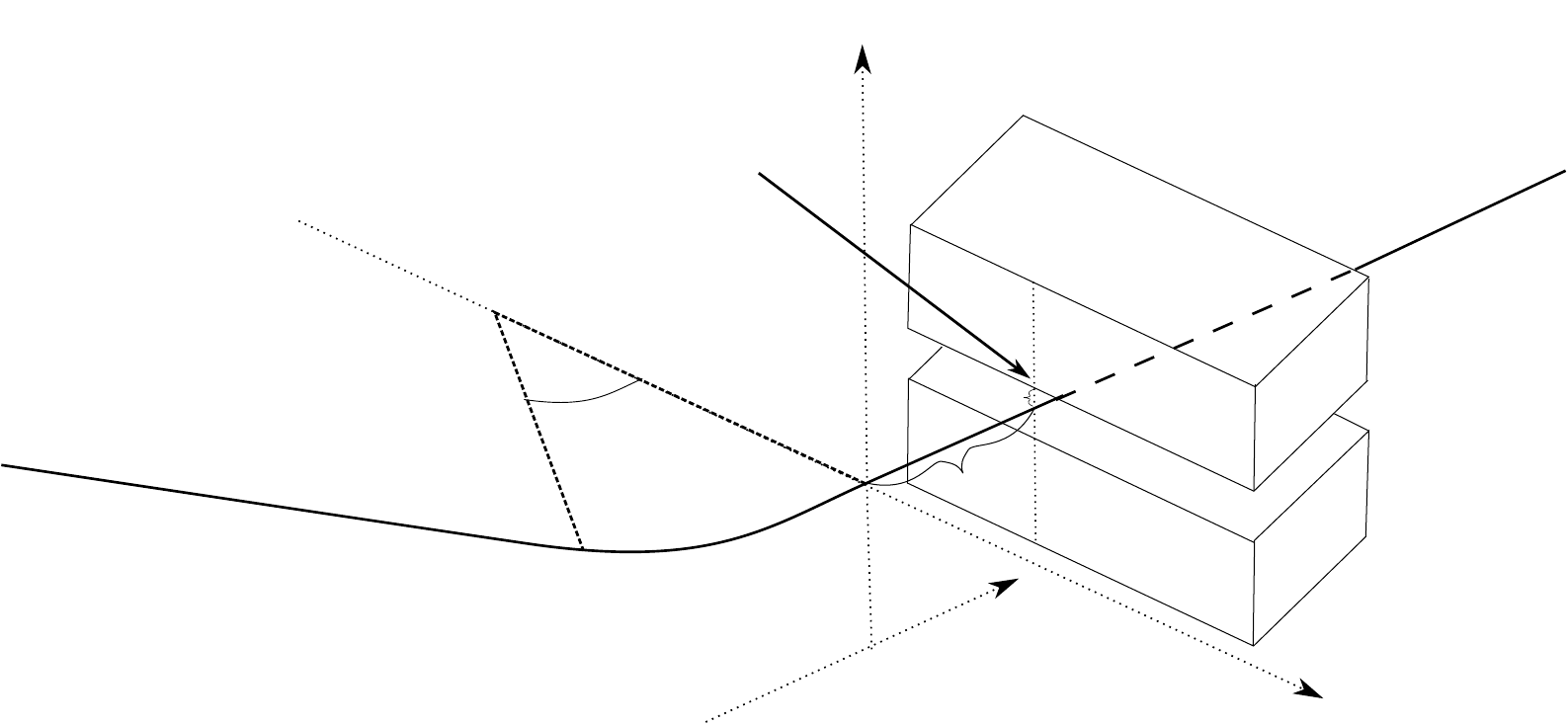}}
%{\color{green}{\graphpaper[1](0,0)(100,50)}}
\put(2,15){\makebox(0,0)[tl]{\rotatebox{-8}{Path of beam $\Big(\!x(\tau),y(\tau),z(\tau)\!\Big)$}}}
\put(65,35){\makebox(0,0)[tl]{\rotatebox{-24}{Collimator}}}
\put(40,40){\makebox(0,0)[tl]{\small Field measurement}}
\put(40,37){\makebox(0,0)[tl]{\small point $\VX$}}
\put(35,17){\makebox(0,0)[tr]{$R$}}
\put(37,20){\makebox(0,0)[tl]{$\Theta$}}
\put(61,16){\makebox(0,0)[t]{$Z$}}
\put(65.1,21.7){\makebox(0,0)[tr]{\footnotesize$h$}}
\put(63,7){\makebox(0,0)[tl]{$z$}}
\put(85,2){\makebox(0,0)[tl]{$x$}}
\put(55,44){\makebox(0,0)[b]{$y$}}
\end{picture}
}
\caption{Suggested path of beam though collimator}
\label{fig_Path}
\end{figure}
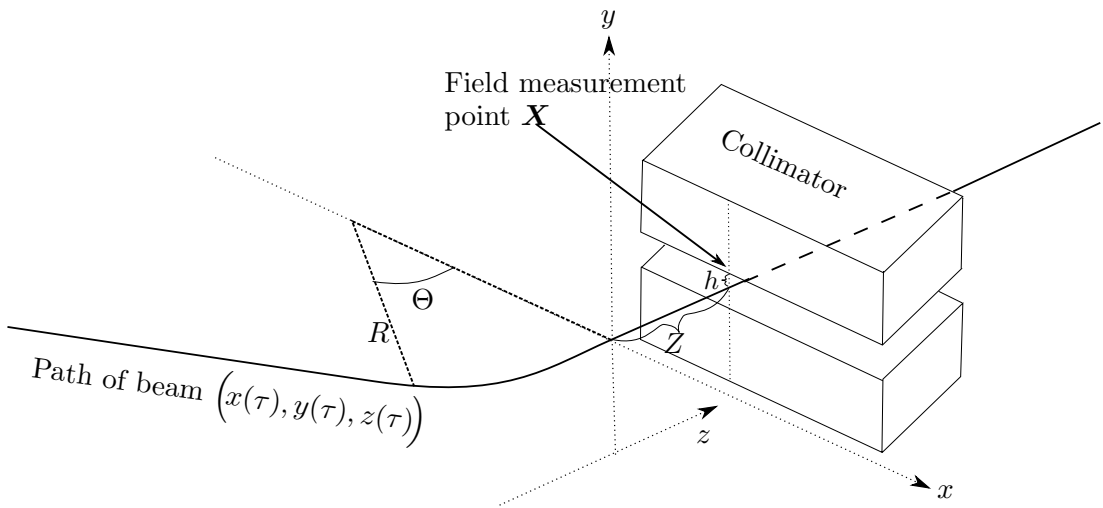

This gives rise to a way to significantly reduce the wakefields
generated when a bunch passes a cavity or collimator. The idea is to bend the beam
slightly before it enters the structure. Most of the Coulomb
field generated by the particle before the bend will continue in a straight line (see FIG. \ref{fig_Path}). By sufficiently
enlarging the beam pipe in this direction the wakefield due to this part of the
field can be neglected. If the distance, $Z$, of the
straight line segment from the terminus of the bend to the centre
of the collimator is sufficiently small, then the resulting pancake field will be
too small to reach the sides of the structure. Of course bending
the beam will generate additional radiation fields, however by judicious choice of
geometry of the beam these can be minimized.

Let $h$ denote half the aperture of the
collimator and let $L$ represent the spatial length of the bunch. The following two scenarios will be considered:
\begin{itemize}
\item Long smooth bunches
where $L>h$ and
any variation in the density of the bunch is over length
scales longer than $h$,
\item Bunches where variation in density is over short length scales less
than about $0.2h$. This includes the case of very short bunches where
$L\ll 0.2 h$.
\end{itemize}

These two scenarios are both applicable to present day machines, where the bunch length depends upon the
specific objectives and engineering considerations of individual projects.
In the following calculation it will be shown that for short bunches, or bunches with large
amounts of micro-bunching, it is possible to make a significant reduction
in wakefields. This is applicable to present day free electron lasers, which employ bunch compressors to produce
very short bunches, for example in LCLS $L/c \approx 0.008$ps. Assuming a collimator of half aperture $h=0.5$mm then in this case $L=0.0048h$. It turns out that electromagnetic fields due to
long smooth bunches may not be reduced significantly. In many present day colliders the bunches are designed to be long and
smooth, however in the future short bunch colliders may be desirable (see TABLE \ref{table1}).
\begin{table}[ht]
\caption{Bunch lengths for some modern colliders and FELs}
\begin{tabular}{|c| c |c|}\hline
Collider & Year of & Bunch length  [ps]\\
 &Commissioning & \\\hline
%LEP, CERN   &  $1989$            & $9000$\\
SLC, SLAC   &  $1989$            & $3$\\
ILC         &  $\geq 2015$       & $1$\\
CLIC        &  $\geq 2025$       & $0.15$\\\hline
Free Electron Laser  &   & Min. bunch length  [ps]\\\hline
FLASH, DESY &  $2005$            & $0.05$\\
LCLS, SLAC  &  $2009$            & $0.008$\\
XFEL, DESY  &  $2014$            & $0.08$\\\hline
\end{tabular}
\label{table1}
\end{table}

Consider a bunch modelled as a one dimensional continuum of point particles where each particle undergoes the
same motion in space but at a different time. This bunch is moving at
a constant speed with relativistic factor $\gamma$. Let $\nu$ label the points
in the bunch, which will be called body points. The profile of the bunch is given by $\rho(\nu)$. Let $\Vx(\tau)$ represent the path of the bunch where $\tau$ is the
proper time. For each body point $\nu$,
%[
\begin{align}
\Vx_\nu(\tau)=\Vx(\tau),\quadand
t_\nu(\tau)=(\tau+\nu)/\gamma.
\label{eqns_path_nu}
\end{align}
%]

In the following all fields are measured at a fixed
point $\VX=(X,Y,Z)$. In FIG. \ref{fig_Path}, $X=0$ and $Y=h$.
 Let $\tauhat(\VX,T,\nu)$ represent the retarded time for the body
point $\nu$ corresponding to the fields measured at $\VX$ at
laboratory time $T$. The retarded time condition is given by
%[
\begin{align}
cT-ct_\nu\big(\tauhat(\VX,T,\nu)\big)=
\norm{\VX-\Vx_\nu\big(\tauhat(\VX,T,\nu)\big)},
\label{eqns_ret_time}
\end{align}
%]
and hence
%[
\begin{align}
\!\!\!\!cT-c\tauhat(\VX,T,\nu)/\gamma-c\nu/\gamma =
\norm{\VX-\Vx\big(\tauhat(\VX,T,\nu)\big)}.
\label{eqns_ret_time_res}
\end{align}
%]
Let $\THat( \nu,\tau,\VX)$ represent the arrival time at $\VX$ of the field
generated by body point $\nu$ at proper time $\tau$. Thus
%[
\begin{align}
c\THat(\nu,\tau,\VX)
&=
ct_\nu(\tau)+\norm{\VX-\Vx_\nu(\tau)}\nonumber\\
&=c(\tau+\nu)/\gamma+\norm{\VX-\Vx(\tau)}.
\label{eqns_def_That}
\end{align}
%]
From (\ref{eqns_ret_time_res}) and (\ref{eqns_def_That})
%[
\begin{align}
cT&=c\big(\tauhat(\VX,T,\nu)+\nu\big)/\gamma+\norm{\VX-\Vx\big(\tauhat(\VX,T,\nu)\big)}\notag\\
&=c\THat(\nu,\tauhat(\VX,T,\nu),\VX).
\label{eqns_tau_T_Invers_a}
\end{align}
%]
Since $\THat$ is increasing and the range of $\tauhat$ is
from $-\infty$ to $+\infty$ it follows that $\THat$ and $\tauhat$ are inverse to each other, yielding (\ref{eqns_tau_T_Invers_a})
and
%[
\begin{align}
\tauhat(\VX,\THat(\nu,\tau,\VX),\nu)=\tau.
\label{eqns_tau_T_Invers_b}
\end{align}
%]
Let us set
%[
\begin{align}
\tauhat_{0}(\VX, T)\!=\!\tauhat(\VX,T,0)
\quadand
\THat_{0}(\tau,\VX)\!=\!\THat(0,\tau,\VX).
\label{eqns_zero}
\end{align}
%]
Then $\THat(\nu,\tau,\VX)$ and $\tauhat(\VX,T,\nu)$ may be written in terms of
$\THat_{0}(\tau,\VX)$ and $\tauhat_{0}(\VX,T)$. From (\ref{eqns_def_That})
%[
\begin{align}
\THat(\nu,\tau,\VX)\!=\!\THat_{0}(\tau,\VX)\!+\!\nu/\gamma.
\label{eqn_THat_nu}
\end{align}
%]
From (\ref{eqns_tau_T_Invers_a}), (\ref{eqns_tau_T_Invers_b}) and
(\ref{eqns_zero}),
%[
\begin{align}
\THat_0(\tauhat_0(\VX,T), \VX)=T
\label{eqns_tau_T_Invers_zero_a}
\end{align}
%]
and
%[
\begin{align}
\tauhat_0(\VX,\THat_0(\tau,\VX))=\tau.
\label{eqns_tau_T_Invers_zero_b}
\end{align}
%]
Substituting (\ref{eqn_THat_nu}) into (\ref{eqns_tau_T_Invers_zero_b}) leads to
%[
\begin{align}
\tauhat_0(\VX,\THat(\nu,\tau,\VX)-\nu/\gamma)=\tau.
\label{eqns_tau_T_Invers_sub}
\end{align}
%]
Substituting $\tau=\tauhat(\VX,T,\nu)$ and using (\ref{eqns_tau_T_Invers_a}) yields
%[
\begin{align}
\tauhat(\VX,T,\nu)\!=\!\tauhat_{0}(\VX, T\!-\!\nu/\gamma).
\label{eqns_tau_nu}
\end{align}
%]

For the body point $\nu$ the Li\'{e}nard-Wiechert electric and magnetic
fields at point $\VX$ and time $T$ are given by substituting
$\tau_R=\tauhat(\VX,T,\nu)$ into (\ref{EB_TX}),
%[
\begin{align*}
&\VE(\!\VX\!,\!T\!,\!\nu\!)\!=\!
\VE_{\text{LW}}\Big(\!\!\VX\!-\!\Vx\big(\!\tauhat(\!\VX\!,\!T\!,\!\nu\!)\!\big),
\Vbeta\big(\!\tauhat(\!\VX\!,\!T\!,\!\nu\!)\!\big),\!
\Va\big(\!\tauhat(\!\VX\!,\!T\!,\!\nu\!)\!\big)\!\Big)
\end{align*}
%]
and likewise for $\VB(\VX,T,\nu)$.
Let $\VEE(\VX,T)$ be the electric field at point $\VX$ and time $T$ due to the body point $\nu=0$ given by
%[
\begin{align*}
\VEE(\VX,T)=
\VE_{\text{LW}}\Big(\!\VX\!-\!\Vx\big(\!\tauhat_0(\!\VX\!,\!T\!)\!\big)\!,\!
\Vbeta\big(\!\tauhat_0(\!\VX\!, \!T\!)\!\big)\!,\!
\Va\big(\!\tauhat_0\!(\!\VX\!,\!T\!)\!\big)\!\Big)
\end{align*}
%]
Using
(\ref{eqns_tau_nu}) it follows
%[
\begin{align*}
\VE(\VX,T,\nu)
&\!=
\VEE(\VX,T-\nu/\gamma).
\end{align*}
%]
The total electric field at the point $\VX$ at time $T$ is given by
%[
\begin{align*}
\VET(\VX,T)
&=
\int\rho(\nu)\VE(\VX,T,\nu)d\nu\\
&=
\int\rho(\nu)\VEE(\VX,T-\nu/\gamma)d\nu
\\&=
\int\gamma\rho\big(\gamma(T-T')\big)\VEE(\VX, T')d T'\\
&=
\int \rhoLab(T-T')\VEE(\VX,T')d T',
\end{align*}
%]
where $T'=T-\nu/\gamma$, and $q\rhoLab(T)=q\gamma\rho(\gamma T)$ is the
charge density as measured in the laboratory frame. Thus the key result
is that the total electric field is given by the
convolution
%[
\begin{align}
\VET(\VX, T)
&=
\int \rhoLab(T-T')\VEE(\VX, T')d T'
\label{E_Tot}.
\end{align}
%]
The above can be repeated for the total magnetic field
$\VB_{\text{Tot}}(\VX, T )$.
Clearly $\VEE(\VX, T')$ will depend on the energy of the beam $\gamma$
and the path of the beam $\Vx(\tau)$. The energy of the
beam is fixed, therefore the only permitted freedom is to alter the position of the
collimator and hence change $\VX$, or to modify the path of the beam.

Consider the path constructed from a straight line followed by an arc of a
circle of radius $R$ followed by another straight
line. Let $\Theta$ denote the angle of arc. The coordinate system is chosen so that the direction of the
second straight line is along the $z$ axis and the arc is in the
$x-z$ plane, finishing at the origin. The trajectory
$\Vx_{\nu}(\tau)=\Vx(\tau)=\big(x(\tau),y(\tau),z(\tau)\big)$ is given in TABLE \ref{path_table}. The point $\VX$ is given as
$\VX = \big(X , Y , Z\big)$ with $Z>0$. Thus the field measured is a function of $\VX$, $\Theta$, $R$ and the time the field arrives, $T$.
\begin{table}[ht]
\caption{The trajectory for a particle in the bunch}
\begin{tabular}{|c|c|c|}\hline
\multicolumn{2}{|c|}{trajectory}&domain\\\hline
&$R(\cos\Theta\!-\!1)\!+\!(\Theta R\!+\!\gamma v\tau)\sin\Theta$&$ -\infty<\tau<-{ R\Theta}/{\gamma v}$\\
$x(\tau)$&$R\Big(\!\cos({\gamma v\tau}/{R})\!-\!1\!\Big)$&$-{R\Theta}/{\gamma v}<\tau<0$\\
&$0$&$ 0<\tau<\infty$\\\hline
$y(\tau)$&$ 0$& $-\infty<\tau<\infty$\\ \hline
&$ -R\sin\Theta\!+\!(\Theta R\!+\!\gamma v\tau)\cos\Theta$& $-\infty<\tau<-{ R\Theta}/{\gamma v}$\\
$z(\tau)$&$R\sin({\gamma v\tau}/{ R})$&$-{ R\Theta}/{\gamma v}<\tau<0$\\
&$\gamma v\tau$&$0<\tau<\infty$\\\hline
\end{tabular}
\label{path_table}
\end{table}

\setlength{\unitlength}{0.7cm}
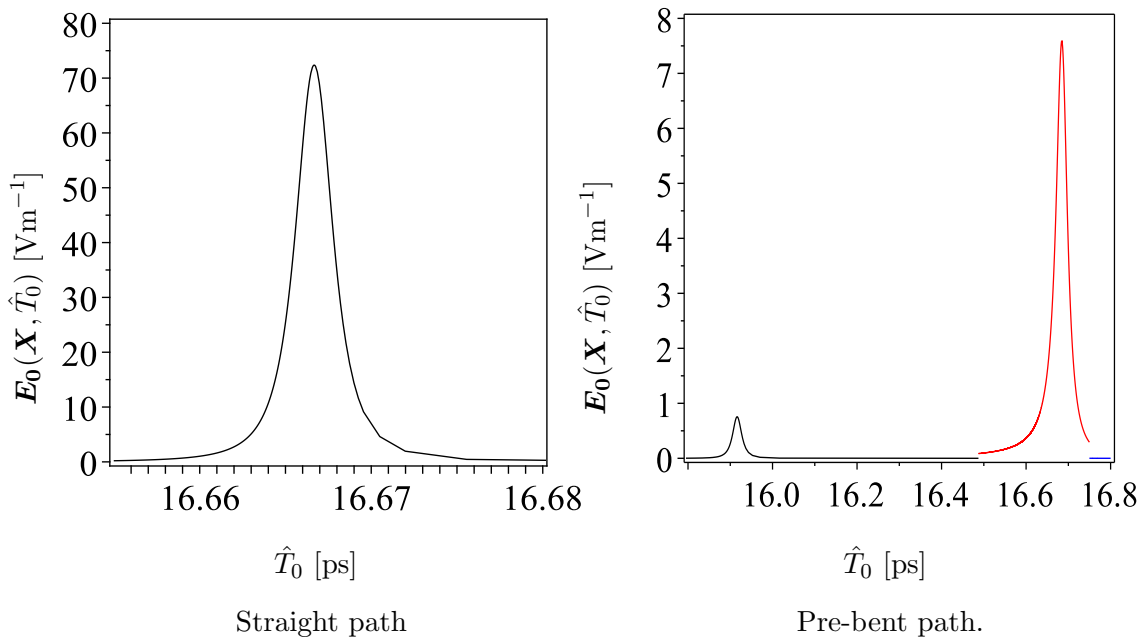
\begin{figure}[ht]
\begin{tabular}{cc}
\begin{picture}(10.5,10.7)
\put(0.2,0.5){
\includegraphics[height=10.7\unitlength]{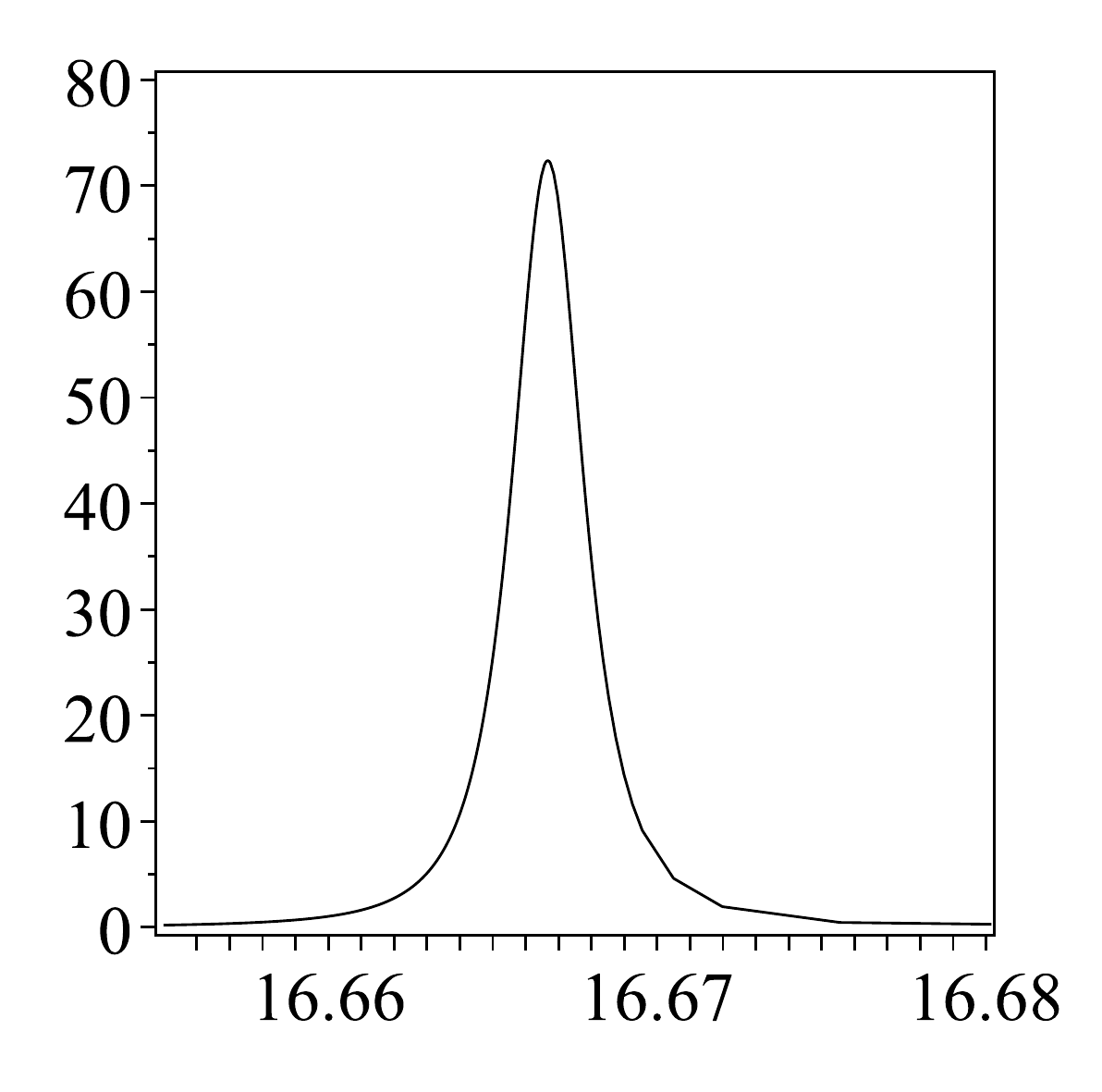}
}
\put(0,3){\rotatebox{90}{\small{ $\VEE(\VX, \THat_0)$ $[\textup{Vm}^{-1}]$}}}
\put(5,0){\small{$\THat_0$ [ps]} }
\end{picture}
&
\begin{picture}(10.5,10.7)
\put(0.5,0.5){
\includegraphics[height=10.7\unitlength]{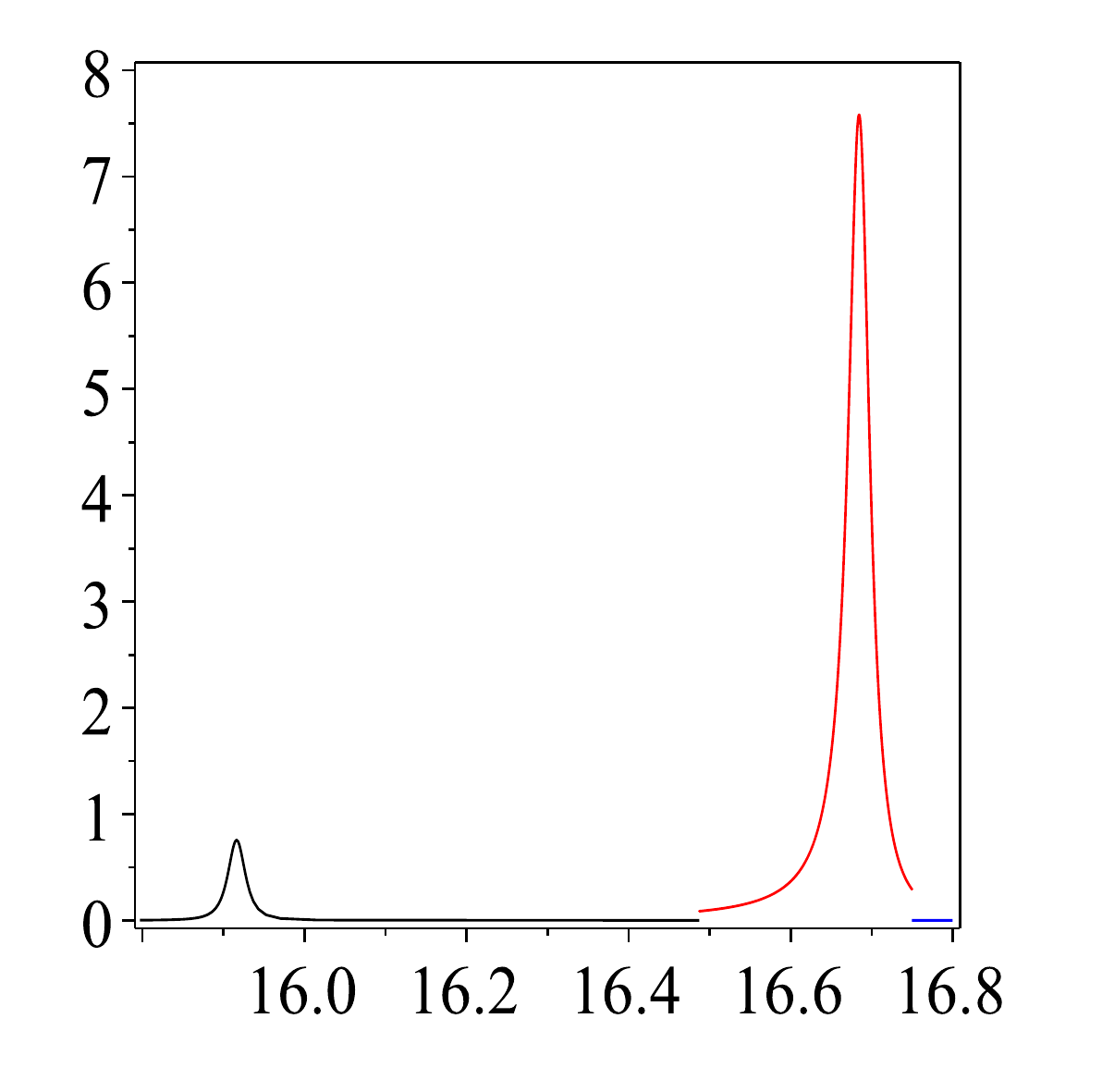}
}
\put(0,3){\rotatebox{90}{\small{$\VEE(\VX, \THat_0)$ $[\textup{Vm}^{-1}]$}}}
\put(5,0){\small{$\THat_0$ [ps]}}
\end{picture}
\\
$\qquad$ Straight path
&
$\qquad$ Pre-bent path.
\end{tabular}
\caption{The electric field strength $||\VEE(\VX, T)||$ at $\VX=(0, h, 10 h)$, with $h=0.5$mm, due to a body point following a
 straight path along the $z$-axis and a body point following the \emph{pre-bent} path given in
TABLE \ref{path_table} with $\Theta=0.13$rad and $R=0.5$m. }
\label{fig_fields}
\end{figure}
Consider the two cases given in FIG. \ref{fig_fields} in which
$\gamma=1000$, $X=0$, $Y=h$ and $Z=10h$. In the straight
line case the peak field is $\approx75$Vm$^{-1}$ and the majority of the field arrives within an interval of
$0.015$ps. In fact it is easy to show that for a straight
line path the peak field increases with $\gamma$ and the width decreases
with $\gamma$ leading to the classic pancake. By contrast for the pre-bent case
 the peak field is significantly reduced to only $\approx7.7$Vm$^{-1}$, however
 the interval over which the field arrives is now $0.35$ps for the right hand peak,
and $0.1$ps for the left hand peak. The reason for these
two peaks is that the left hand peak is the coulomb field due to
the first straight line segment, whereas the second peak is due to the
radiation from the circular part of the beam path. The discontinuity is a result of
 the discontinuity in acceleration for this trajectory. Repeating the
calculation with higher $\gamma$-factors doesn't significantly change the
height or shape of the second peak.

If the bunch is long and smooth, i.e. longer than the collimator
aperture, so that there is no significant change in $\rhoLab$ over the
width of $\VEE(\VX,T')$, then $\VEE(\VX, T')$ may be crudely regarded as a
$\delta$-function and $\VET(\VX,T)$ is given  by
%[
\begin{align}
\VET(\VX,T)
&\approx
\rhoLab(T)
\int \VEE(\VX,T')d T'.
\label{E_Tot_smooth}
\end{align}
%]
Integration of $\VEE(\VX,T')$ for the straight and pre-bent trajectories
 reveals that
 %[
\begin{align}
||\VET(\VX, T)||\approx\frac{q}{2\pi\epsilon_0 c}\frac{\rhoLab(T)}{||\VX||}
\label{E_Tot_smooth_res}
\end{align}
%]
This value of $\VET$ is independent of $R$ and
$\Theta$ for all paths where $R$ is large compared to $L$. To see why
this is the case consider our one dimensional beam of particles as a
continuous flow of charge, similar to a line charge in a wire but
without the background ions. The fields due to this flow may be
calculated using the Biot-Savart law. Since $h\ll R$ the field is
dominated by the nearby current and hence no variation of $R$,
$\Theta$ or $Z$ will alter the fields.
\begin{table}[ht]
\caption{Peak field strength for different sized bunches with h=0.5mm.}
\begin{tabular}{| @{\ \ }l @{\ \ }| @{\ \ } l@{\  \ }| @{\ \ }l@{\ \ }|@{\ \ }l@{\ \ }|}\hline
\multicolumn{2}{|c|@{\ \ }}{Bunch Length} &\multicolumn{2}{c|}{Peak $||\VET(\VX,T)||$ [Vm$^{-1}$]}\\\hline
 \multicolumn{1}{|c|@{\ \ }}{L[h]} & \multicolumn{1}{c|@{\ \ }}{(L/c)[ps]} &  \multicolumn{1}{c|@{\ \ }}{straight}&  \multicolumn{1}{c|}{pre-bent} \\\hline
 $1.8 \times 10^1   $   &  $3.00\times 10^0  $ &  $1.97 \times 10^{-1}$  &  $1.97 \times 10^{-1}$\\
 $6.00 \times 10^{-1}$   &  $1.00 \times 10^0 $ &  $5.91 \times 10^{-1}$  &  $5.89 \times 10^{-1}$\\
 $9.00 \times 10^{-2}$   &  $1.50 \times 10^{-1}$ &  $3.93 \times 10^0   $  &  $3.48 \times 10^0   $\\
 $4.80 \times 10^{-2}$   &  $8.00 \times 10^{-2} $ &  $7.33 \times 10^0   $  &  $5.27 \times 10^0   $\\
 $3.00\times 10^{-2} $   &  $5.00 \times 10^{-2} $ &  $1.16 \times 10^1   $  &  $6.36\times 10^0    $\\
 $4.80 \times 10^{-3}$   &  $8.00\times 10^{-3}  $ &  $5.12 \times 10^1   $  &  $7.53 \times 10^0   $\\\hline
\end{tabular}
\label{table3}
\end{table}

If the beam has bunches of length $L\lesssim 0.05 h$ then it follows from
(\ref{E_Tot}) and FIG. \ref{fig_fields} that a considerable reduction in fields is possible. It
is straightforward to calculate $\VET(\VX, T)$ numerically using a
Gaussian particle distribution $\rho_{\text{Lab}}$ for the two cases
in FIG. \ref{fig_fields} (see TABLE \ref{table3}). If $\rho_{\text{Lab}}$ has full width
at half maximum $L/c=0.008ps$ with corresponding bunch length $L=0.0048 h$, then the peak value for the total electric field in the straight line case is given by $\approx 51.2
\textup{Vm}^{-1}$. By contrast, in the pre-bent case the peak value
for the total electric field is $\approx 7.5\textup{Vm}^{-1}$,
giving an approximate factor of 7 reduction in field. This is approaching
the maximal factor of 10 improvement one can achieve with $\gamma=1000$, which occurs when
the bunch length is small enough that the convolution gives the peak values for the fields in FIG. \ref{fig_fields}. With
higher energies and shorter bunch lengths the radiation peak remains unchanged, whereas the electric field for the
straight path grows linearly with $\gamma$. Thus even greater
improvements can be made.

\begin{figure}[ht]
\begin{center}
\setlength{\unitlength}{0.04\textwidth}
\begin{picture}(10,10)
\put(0,0){\rotatebox{90}{\includegraphics[width=10\unitlength]{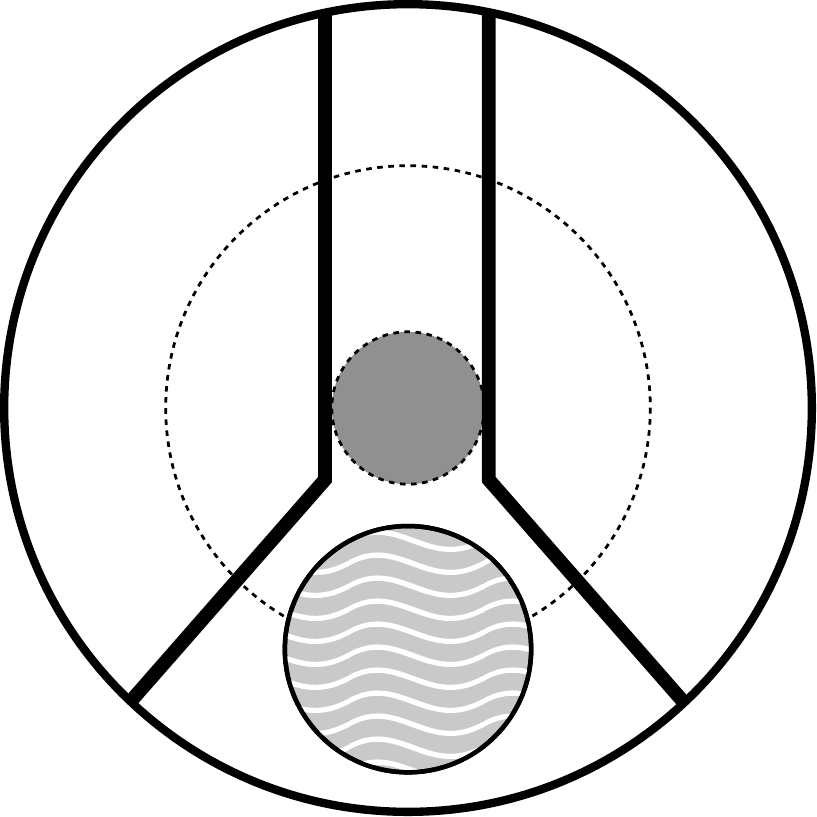}}}
%\graphpaper[1](0,0)(10,10)
\put(-0.,7.6){\footnotesize\rotatebox{45}{beam Pipe}}
\put(2.2,3){\footnotesize\rotatebox{-45}{halo}}
\put(4.5,4.8){\footnotesize\rotatebox{0}{beam}}
\put(1,6.2){\footnotesize\rotatebox{0}{collimator}}
\put(7.7,5){\footnotesize{High}}
\put(7.7,4.3){\footnotesize{Fields}}
\end{picture}
\end{center}
\vspace{-2em}
\caption{Modified collimator in the plane transverse to the path of the beam.}
\label{fig_Modified_coll}
\end{figure}
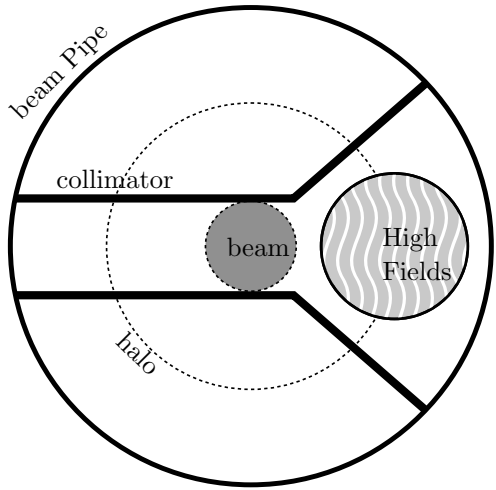

In the above calculation the specific field point
$\VX=(0, h, 10h)$ is chosen and the peak electric fields are minimized
for this particular point on the collimator. If instead the point is
displaced in the positive $x$ direction (FIG \ref{fig_Path}), then a
significant increase in field strength is observed. This increase results from
both the Coulomb field from the straight section of the path before
the arc and the radiation from the circular part of the path. It will
be necessary to alter the shape of the collimator to avoid these high
fields interacting with the material in the collimator. This need not
affect the efficacy of the collimator to remove the halo, for example
see FIG \ref{fig_Modified_coll}. The optimum design of the beam path, beam
tube and collimator shape, for particular machines will require a
combination of analytic, numerical and experimental research.
Clearly long tapers will reduce the advantage gained by bending the beam
since it will give time for the pancake to form. However it may be
advantageous to use a short taper.

The placing of an additional bending dipole just before a collimator
would inevitably cause unwanted losses in beam energy due to radiation
loss. However all accelerators, even Linacs, already have to bend the
beam using dipoles in certain places. Therefore it seems natural to
place a collimator directly after a bending magnet in order not to
lose any more beam energy through radiation loss.

The authors acknowledge support from the Cockcroft Institute (STFC
ST/G008248/1) and the Alpha X project Strathclyde University. They are
grateful to Dr D. A. Burton Lancaster University, and Dr A. Noble
Strathclyde University for valuable discussions.

%%%%%%%%%%%%%%%%%%%%%%%%%%%%%%%%%%%%%%%%%%%%%%%%%%%%%%%%%%%%%%%%%%%%%%
\bibliographystyle{unsrt}
%\bibliography{Sus}
%%%%%%%%%%%%%%%%%%%%%%%%%%%%%%%%%%%%%%%%%%%%%%%%%%%%%%%%%%%%%%%%%%%%%%

\bibliographystyle{unsrt}
\bibliography{bendingbeams}

\begin{thebibliography}{10}

\bibitem{Yokoya90}
K.~Yokoya.
\newblock Impedence of slowly tapered structures.
\newblock {\em CERN Report}, (SL-90-88-AP), 1990.

\bibitem{Warnock93}
R.~L. Warnock.
\newblock An intergo-algebraic equation for high frequency wake fields in a
  tube with smoothly varying radius.
\newblock {\em SLAC Report}, (SLAC-PUB-6038), 1993.

\bibitem{Stupakov96}
G.~V. Stupakov.
\newblock Geometrical wake of a smooth flat collimator.
\newblock {\em SLAC Report}, (SLAC-PUB-7167), 1996.

\bibitem{Stupakov01}
G.~V. Stupakov.
\newblock Impedance of small-angle collimators in high-frequency limit.
\newblock {\em SLAC Report}, (SLAC-PUB-8857), 2001.

\bibitem{Stupakov07}
G.~Stupakov.
\newblock Low frequency impedance of tapered transitions with arbitrary cross
  sections.
\newblock {\em Phys. Rev. ST Accel. Beams}, 10(9):094401, Sep 2007.

\bibitem{Bane07}
K.~L.~F. Bane, G.~Stupakov, and I.~Zagorodnov.
\newblock Impedance calculations of nonaxisymmetric transitions using the
  optical approximation.
\newblock {\em Phys. Rev. ST Accel. Beams}, 10(7):074401, Jul 2007.

\bibitem{Bane10}
G.~Stupakov, K.~L.~F. Bane, and I.~Zagorodnov.
\newblock Impedance scaling for small angle transitions.
\newblock {\em Phys. Rev. ST Accel. Beams}, 14(1):014402, Jan 2011.

\bibitem{Podobedov06}
B.~Podobedov and S.~Krinsky.
\newblock Transverse impedance of axially symmetric tapered structures.
\newblock {\em Phys. Rev. ST Accel. Beams}, 9(5):054401, May 2006.

\bibitem{Podobedov07}
B.~Podobedov and S.~Krinsky.
\newblock Transverse impedance of tapered transitions with elliptical cross
  section.
\newblock {\em Phys. Rev. ST Accel. Beams}, 10(7):074402, Jul 2007.

\bibitem{Smith11}
J.~D.~A. Smith.
\newblock {\em Calculations of Collimator Wakefield}.
\newblock PhD thesis, Lancaster University, UK, 2011.

\bibitem{Jackson99}
J~D Jackson.
\newblock {\em Classical Electrodynamics (3rd Edition)}.
\newblock Wiley, 1999.

\end{thebibliography}

\end{document}